\documentclass{appolb}
\usepackage{graphicx}
\usepackage{amsmath}

\def\vec#1{\ensuremath{\mathchoice
                     {\mbox{\boldmath$\displaystyle#1$}}
                     {\mbox{\boldmath$\textstyle#1$}}
                     {\mbox{\boldmath$\scriptstyle#1$}}
                     {\mbox{\boldmath$\scriptscriptstyle#1$}}}}

\begin{document}
%
\title{A  local reduction  of the Dirac equation applied to the study 
of quark interactions
}
\author{M. De Sanctis \footnote{mdesanctis@unal.edu.co}
\address{Universidad Nacional de Colombia, Bogot\'a, Colombia }
\\
}
\maketitle
\begin{abstract}
A general procedure of local reduction  for the Dirac equation 
 is introduced to study one- and n-body interacting systems.
In the one-body case we show that the reduction allows for an approximate
solution of the Dirac equation, correlating the upper and the lower components
of the wave function.
The two-body case is studied in more detail. 
We show that the method prevents from introducing spurious,
unphysical states. 
The reduction is also applied to another relativistic equation.
Finally, the method is used to construct a specific model in order to
 study the Charmonium spectrum.
\end{abstract}
\section{Introduction}
Relativistic wave equations for bound systems represent a relevant,
controversial and extremely extended area of investigation in theoretical physics.
The aim of this work is to propose a three-dimensional reduction of the Dirac 
and Breit equations in order
to describe in a relativistic way, avoiding some known inconsistencies, 
the dynamics of spin 1/2 particles bound states.\\
The article is organized as follows:
in the remainder of the introduction, in Subsect.  \ref{contextaim} we contextualize
 the present study in the field of the relativistic equations and define the aim of the work; 
in Subsect. \ref{symbnot}
we introduce the symbols and the notation of the article;
then, in Sect. \ref{diracappr} we study an approximate formal method of solution 
of the one-body Dirac equation;
taking advantage of the obtained results, in Sect. \ref{redpro} we discuss reduction procedure
starting from the one-body case;
in Sect. \ref{twobody} we generalize the reduction procedure to the two-body case,
that will be analyzed in more detail, and to the n-body case that will be 
introduced at formal level;
in Sect. \ref{applmw} we apply our reduction procedure to the modified Dirac equation proposed by Mandezweig and Wallace;
in Sect. \ref{charm} we introduce, as an example, an interaction model for the study of the Charmonium spectrum;
the numerical method of solution  for the equation is briefly discussed 
in Sect. \ref{solution};
 in Sect. \ref{charmres} we discuss the results and make a comparison
 with the experimental data for the Charmonium spectrum;
finally, some conclusions are drawn in Sect. \ref{conclusions}.
Appendix \ref{tr2bwe} is devoted to analyze the technical details of some 
three-dimensional relativistic equations, particularly relevant for this work.
The reduction of the one- and two-body interaction  is analyzed in the
Appendixes \ref{reduct1} and \ref{reduct2}, respectively.

\subsection{Context and aim of the work}\label{contextaim} 
In this subsection, before defininig the objective of the work, 
we try to contextualize the 
content of the present paper in the framework of the 
 three-dimensional relativistic wave equations (TDRWEs),
with no attempt to cover the whole subject.
For a  concise description of the technical details about the TDRWEs
 related to this work,
the interested reader is referred to Appendix \ref{tr2bwe}.\\
We start by recalling that
Dirac equation represents the basic element for the study of all
the field theories, as QED, electroweak theory and QCD.
Historically, that equation, considered as a \textit{one-body}
relativistic equation for a \textit{bound} electron in an external potential,
has been successfully used to determine
the fine structure effects of the Hydrogen atom spectrum. \\
On the other hand, the study of  relativistic equations for two- and
n-body \textit{bound} systems 
(that is very  relevant for atomic, nuclear and, particularly, for  hadronic physics)
is much more complex and many different strategies 
have been proposed. \\
In principle,
it is possible to construct the Hamiltonian for two- and n-body systems
as a straightforward sum of the one-body Dirac Hamiltonians. 
This procedure gives rise to  the Breit equation \cite{breit1,breit2,breit3}, that will be
conventionally  denoted in the present paper as \textit{Dirac-like equation} (DLE)
in order to emphasize its relationship with the original Dirac theory.\\
%
The DLE is a TDRWE in Hamiltonian form.
A specific, relevant advantage of this equation consists 
in its \textit{full locality},
if a local interaction is taken.\\
In recent years the derivation of the DLE  has been formally revised 
 considering direct interactions
between spin $1/2$ particles \cite{louism}.
Furthermore, this equation
 has been  deeply analyzed and successfully used to calculate finite size
\textit{perturbative corrections} in the Hydrogen atom \cite{garciakelkar} and in mesonic atoms 
\cite{kelkarnow}.\\
A semi-analytic study has been performed for some bound states with a 
Coulomb potential \cite{kasa}. 
A calculation of the spectra of quarkonia has been also developed \cite{tsibi}.\\
However, for that equation 
a relevant formal difficulty has been found: the so-called \textit{continuum dissolution problem} (CDP)  \cite{such}.
Essentially, it is related to the presence of spurious null mass solutions
for the case of non-interacting particles; in more detail, 
considering a two-body system,
it is possible to have a positive energy solution for one particle and 
a negative energy solution  for the other particle;
in consequence, in the rest frame, a zero total mass is obtained.
Equivalently, these null mass free solutions correspond to 
unphysical  poles in the three-dimensional Green function, 
as shown in Eq. (\ref{g2dl}).  
The CDP is related to the difficulty  of treating
the negative energy states generated  by the  one-body  Dirac terms 
of the Hamiltonian.\\
As a consequence, in \textit{nonperturbative} calculations, 
the reliability of the solutions of the DLE is strongly questioned,
while it can be safely used in the perturbative ones.\\
Many other different methods have been  followed to study relativistic bound states.
Among them, we only quote the \textit{Dirac's constraint dynamics} \cite{cra1,cra2,cra3}
and the
\textit{relativistic path integral Hamiltonian approach} \cite{sim1,sim2,sim3}.\\
We now discuss some models,
more strictly related to the present work, that  have been derived from 
the Bethe-Salpeter  equation (BSE) \cite{sabe,bethes}; for this equation
an extensive didactic exposition can be found in Ref. \cite{bsed}. \\
The BSE is an explicitly covariant four-dimensional formalism that, 
\textit{in principle},
allows to  sum up the infinite series of all the Feynman graphs for two
interacting particles, reproducing completely the dynamics of the bound system. 
However, this procedure would require to introduce in the interaction 
kernel \textit{all} the corresponding \textit{irreducible} Feynman graphs.
Unfortunately, this task cannot be accomplished: only the \textit{tree-level}
boson exchange graph is usually considered for the kernel.
In this way the BSE could only reproduce the series of the 
\textit{ladder} graphs.
But at this point also another problem is found: the tree-level  
boson exchange graph, due to its singularities, gives rise to 
abnormal (unphysical) solutions \cite{lucha}; for this reason one is forced to
assume an \textit{instantaneous} tree-level interaction.
With this assumption, the BSE is reduced to the three-dimensional 
Salpeter instantaneous   equation (SIE) \cite{bsed,salp}.
(In this concern, we recall that for the electromagnetic interaction in the Coulomb gauge,
the Coulomb term is instantaneous.)
The SIE is, in any case, a \textit{nonlocal}  equation that is practically 
written as an integral equation by means of the Green function 
that propagates the $++$ and the $--$ states 
\textit{but not} the $+-~$ and $-+$ states
that are excluded from the model.
For this reason the Green function of this model is not invertible.
The technical details  about this point are given in Appendix \ref{tr2bwe}; in particular, see Eq. (\ref{g2si}) .
Incidentally, we recall that
a comparison between the numerical solutions of the SIE  and of
the DLE has been performed, finding for the DLE equation 
unphysical effects related to the CDP \cite{chang}.\\ 
Many efforts have been devoted to improve the SIE trying to incorporate,
with some approximation, the \textit{crossed} graphs in order  to go beyond the 
\textit{ladder} approximation for the full series of the equation.\\
This objective has been achieved in part by putting \textit{on-shell}
one  fermion, in the so-called \textit{Relativistic Spectator Formalism},
originally developed for nuclear systems and also applied to quark bound states;
see, for example Refs. \cite{grossa,grossb,grossc,cst1}. \\
In another approach, that is the Mandelzweig and Wallace equation (MWE) 
\cite{mwa,mwb,mwc},
 the crossed graphs are taken into account, in the eikonal approximation,
by means of a suitable definition of the Green function.
In this way, the Green function is that of the SIE, \textit{plus}
the contributions of the $+- $ and $-+$ states, as shown in Eq. (\ref{g2mw}).
As a result one obtains an \textit{invertible} Green function that finally 
gives rise to a modified Dirac equation for the bound state. 
This equation has not the form of an eigenvalue equation for the total energy 
of the system; the non-interacting term depends on nonlocal operators
but the \textit{interaction term has a local form}.
Due to its structure, the  MWE is free from the CDP and represents an interesting improvement 
with respect to the SIE.
However,
when applied to an effective gluon exchange interaction,
it should be carefully reexamined considering in particular
the noncommutativity of the interaction vertices for the crossed graphs.\\
The MWE has been also applied to study relativistic corrections for
few-body nuclear systems,
taking into account, in that case,  the non-Abelian character of the one-pion exchange 
interaction \cite{dspa,dspb,dspc}.\\
Given the structure of the MWE
and, in particular, due to the 
locality of the interaction term, we shall apply also to this model 
the reduction procedure developed in the present work.\\ 
A common problem of the DLE and MWE equations is the lack of explicit
relativistic covariance. This problem is standardly solved by defining
in a covariant way the variables of the center of mass, 
where these equations are originally derived \cite{mwb,dspb,mosh}.\\
Finally, we note that the contribution of the $++$ states to the Green function 
is \textit{the same} 
for the  DLE, SIE and MWE.\\
For this reason, and also
considering the dynamical uncertainties discussed above and the
difficulties of the numerical solutions,  
a possible starting point for a relativistic study of the bound systems 
consists in projecting  any TRWE, that is the DLE, SIE or MWE,
\textit{only} onto the positive energy states, excluding completely the
$+-$, $-+$ and $--$ states.
In this way, the equation 
shown in Eqs. (\ref{eqplusplus}) and (\ref{eqplusplusint}) has been obtained.
 We denote this equation as
\textit{positive energy state equation} (PESE).
Moreover, it is possible to introduce into this equation some retardation contributions
without formal difficulties.
This equation  has been used to study the spectra of heavy
quarkonia \cite{mdsf,mdsfs}.
\vskip 1.0 truecm
\noindent
The previous discussion shows that the problem of the relativistic
equations for bound states is still an open issue, with different levels of complexity.\\
From the numerical point of view,
in the two-body case, for the DLE and MWE, one has to solve
a coupled equation for \textit{four} two-component spinors.
For the DLE  one has an eigenvalue equation with standard differential 
operators in the coordinate space.
For the MWE, the non-interacting term is energy-dependent and nonlocal;
in consequence,  a specific strategy should be studied. 
For the SIE, due to its nonlocal form, one has to solve  a coupled integral equation for \textit{two} two-body spinors,
corresponding to the $++$ and $--$ components of the wave function.
In the case of Eq. (\ref{eqplusplusint}) for $++$ states only, one has an integral equation for \textit{one} two-body  spinor.\\
From the dynamical point of view, we note that, in any case, 
starting from a hypothetical exact theory, 
many (and not completely justified) approximations are required
to define a specific model.\\
In particular, assuming that the BSE represents the correct starting point,
one has to take into account that
the crossed graphs are not included or approximately 
included in the SIE and MWE, respectively.
On the other hand,
the DLE can be considered an equation  
 based on \textit{first principles} but  
the unphysical singularities of the Green function must be removed.\\
This situation has motivated the development of the present study.\\
In particular, the aim is to find 
a \textit{reduction} of the DLE
equation by establishing a relationship, or \textit{correlation},
  between the lower and upper components of the Dirac spinors of each interacting fermion. 
This correlation has the same structure of the solutions 
of the one-body Dirac equation 
in the ``spin-symmetry" case \cite{spinsym1,spinsym2}.
	We recall that a simplified Dirac-Coulomb equation was proposed for
	atomic systems \cite{pest}.
	A Dirac harmonic oscillator shell model with spin-symmetry
	was also used to study quark-antiquark spectroscopy \cite{bhagh}.\\
The reduction of this work is mainly oriented to the study of 
\textit{few-body hadronic systems}
and  avoids from the beginning the CDP.
An energy-dependent, three-dimensional, completely 
\textit{local} equation is obtained; consequently
a relatively simple numerical solution  is achievable in the coordinate
space.
For the two-body case, one has to handle only \textit{one} two-body spinor.\\
Our reduction can be considered equivalent to the standard PESE
when applied to  the scattering of on-shell particles but  
includes, for bound states, some contributions of the negative energy states.
More details are given  in Eqs. (\ref{amppos}), (\ref{ampneg}).

\noindent
After studying the reduction procedure for the DLE,
we also apply it to the MWE.
In this case, for the reduced noninteracting term  of the equation,
we obtain a nonlocal operator
but, for the reduced interaction operator, 
we have
the same local form obtained for the DLE.
In consequence, it is possible to use for the variational solution,
the harmonic oscillator (HO) basis, that admits an analytic Fourier transform.
We shall use the momentum space HO wave functions for the noninteracting term,
while the coordinate space wave functions will be used for he interaction term,
as discussed in Sect. \ref{solution}.

%

\noindent
As an example of application,
we use our reduced equation to  study the Charmonium spectrum
by means of a standard interaction given by a vector and a scalar term.
The results given in Sect. \ref{charmres} show that a good quality reproduction of the spectrum can be obtained.

\vskip 0.5 truecm
\noindent
\subsection{Symbols and Notation}\label{symbnot}
In the present work we use 
the gamma matrices $\gamma^\mu$  in the standard representation.
Given that we shall employ the \textit{Hamiltonian formalism} 
for the Dirac equation, 
we also introduce $\beta= \gamma^0$ and
the matrices 
$\gamma^0 \gamma^\mu=(\mathcal{I}, \vec \alpha)$ where $\mathcal{I}$
represents the identity matrix in the $4 \times 4$ Dirac space.\\
In the two- and n-body cases, for the matrices and the operators 
a \textit{particle} lower index $i=1,2,...n$ is introduced;
but for the one-body case the index $1$ is omitted.\\
The following shorthand notation is used:
$O_i=O(m_i,\vec p_i;\vec \alpha_i,....)$
where the generic one-body operator 
$O(m,\vec p;\vec \alpha,....)$
 is calculated for the $i$-th particle.\\
An operator specifically introduced for a n-body system will
be denoted by the subscript $(n)$, with parentheses.\\
The letter $\Psi$ denotes the complete Dirac wave functions, for the n-body system.
The letter $\Phi$ is used for the  spinorial (reduced) wave functions.\\
For the general equations, 
we use the \textit{bra-ket} Dirac notation $|\Psi>$, $|\Phi>$.\\
The interaction operators $W_{(n)}$ are referred to 
the Hamiltonian formalism,   that is: 
$\Psi^\dag W_{(n)} \Psi= \overline \Psi \gamma_1^0 \cdot \cdot \cdot \gamma_n^0 W_{(n)} \Psi$.\\
The \textit{reduced} operators derived in the present work will be denoted 
by a \textit{hat}.\\
The spin indices  will be generally omitted.
Only in Sect. \ref{solution} the spin quantum numbers are explicitly indicated
for the variational wave functions.\\
Finally, throughout the work we use the so-called natural units, that is $\hbar=c=1$.\\
\vskip 1.0 truecm

\section{The Dirac equation and its approximate solution}\label{diracappr}
In order to introduce the reduction procedure,
we previously analyze a solution method for the Dirac equation.
Then,   the reduction procedure of the one-body case will be 
studied in Sect. \ref{redpro} and  then generalized 
to two- and  n-body DLE and to MWE.

\noindent
We write the Dirac equation in the Hamiltonian  form:
\begin{equation}\label{dirac1}
(H^{free} -E  +W_{(1)})|\Psi>=0
\end{equation}
where $H^{free}$ represents the standard one-body, free, Dirac Hamiltonian,
that is
\begin{equation}\label{free1}
H^{free}=  H^{free}(m;\vec p, \vec \alpha, \beta) =
\vec \alpha \cdot \vec p +\beta m
\end{equation}
with the Dirac matrices recalled in Subsect. \ref{symbnot};
$\vec p$ and $m $ repectively represent the particle momentum and mass;
furthermore, in Eq. (\ref{dirac1})  $E$ is the energy eigenvalue and 
$W_{(1)}$ represents the one-body interaction with an external field. 
Finally,    $|\Psi>$ represents the four-component Dirac spinor
that in the coordinate representation reads  $\Psi(\vec r)=<\vec r| \Psi>$. \\
In view of the formal development of the work, we also introduce here the 
Dirac  operator, in the form:
\begin{equation}\label{dirop}
D=D(m,E;\vec p, \vec \alpha, \beta)=H^{free}-E~.
\end{equation}
We take, for the following introductory discussion, a specific interaction with
a scalar field $V_s(\vec r)$ and the time component
of a vector field $V^0_v(\vec r)$; in this way the one-body interaction has the form:
\begin{equation}\label{intg1}
W_{(1)}=\beta V_s(\vec r) + V^0_v(\vec r)~.
\end{equation}
We split the four-component one-body Dirac spinor  into two  two-component spinors,
for the \textit{upper} 
and  the \textit{lower}  
components:
\begin{equation}
|\Psi>=
\begin{pmatrix} |\Phi_U >\\
                |\Phi_L >
\end{pmatrix}~.
\end{equation}								
For the interaction fields we introduce
the shorthand notation
\begin{equation}\label{vpm}
\begin{split}
V_U(\vec r)= V^0_v(\vec r)+ V_s(\vec r)~~\\
V_L(\vec r)= V^0_v(\vec r)- V_s(\vec r) ~.
\end{split}
\end{equation}
With this notation, the total interaction of Eq. (\ref{intg1}) can be written as:
\begin{equation}\label{intgg1}
W_{(1)}= {\frac 1 2}   \beta [V_U(\vec r) -V_L(\vec r)] +
   {\frac 1 2}         [V_U( \vec r) +V_L(\vec r)]    ~.              
\end{equation}
%
\noindent
In this way the Dirac equation (\ref{dirac1}) can be conveniently written 
as a matrix equation in the form:

\begin{equation}\label{dirac1matr}
\begin{pmatrix}
m -E +V_U(\vec r) & \vec \sigma \cdot \vec p \\
\vec \sigma \cdot \vec p &  -(m+E) +V_L(\vec r)
\end{pmatrix}
\begin{pmatrix} 
|\Phi_U> \\
|\Phi_L> 
\end{pmatrix}=0
\end{equation}
that
 represents a \textit{coupled equation} for
the two spinors $|\Phi_U>$ and $|\Phi_L>$ .\\
Assuming that the quantity
$ m +E-V_L(\vec r)$
is nonvanishing,
one can express $|\Phi_L>$
by means of $|\Phi_U>$ in the form:
\begin{equation}\label{psilfpsiu}
|\Phi_L>= [m+E-V_L(\vec r)]^{-1} \vec \sigma \cdot \vec p 
~|\Phi_U>
\end{equation}
In this way, one \textit{correlates} exactly the upper and the lower components 
of the Dirac state $|\Psi>$.\\
Then, the equation for $|\Phi_U>$ can be written 
\textit{exactly} as:
\begin{equation}\label{psiuexact} 
\left[m -E +V_U(\vec r) +
\vec \sigma \cdot \vec p 
 (m +E-V_L(\vec r))^{-1}
\vec \sigma \cdot \vec p 
\right] |\Phi_U> =0~.
\end{equation}
The last term of this equation, for a central interaction 
 $V_L=V_L(r)$, can be rewritten by using the
transformation given in Eq. (\ref{simpl1}) of Appendix \ref{reduct1}.
\vskip 0.5 truecm
\noindent
In order to introduce our reduction technique, 
we factorize the constant factor $(m+E)^{-1}$. 
With standard algebra we write :
\begin{equation}\label{algebex}
[m+E-V_L(\vec r)]^{-1}=
[1+ B(E;\vec r)] \cdot{\frac {1} {m+E}}
\end{equation}
with
\begin{equation}\label{ger}
B(E; \vec r)= F(E; \vec r)\cdot
{\frac {1} {1- {F(E;\vec r)} } }
\end{equation}
and
\begin{equation}\label{fer}
F(E;\vec r)={\frac {V_L(\vec r) } {m+E} }
\end{equation}
Replacing Eq. (\ref{algebex})
in Eq. (\ref{psiuexact}),
one obtains:

\begin{equation}\label{psiuuexact}
\left[m -E +
{\frac {{\vec p}^2} {m+E} }  +V_U(\vec r) +
{\frac { 1 } {m+E} }
\vec \sigma \cdot \vec p 
 B(E;\vec r)
\vec \sigma \cdot \vec p 
\right] |\Phi_U> =0
\end{equation}
where one has to remenber the definitions of the Eqs. (\ref{ger}) and (\ref{fer})
for $ B(E; \vec r)$. 
Eq. (\ref{psiuuexact})
is  an energy-dependent, still \textit{exact} equation for $|\Phi_U>$;
 $|\Phi_L>$ can be reconstructed by means of Eq. (\ref{psilfpsiu}).
With respect to Eq. (\ref{psiuexact} )
the previous transformations have allowed to isolate
the energy-dependent \textit{pseudo-kinetic} term, that is:
\begin{equation}\label{tek} 
T_K(E)={\frac {{\vec p}^2} {m+E} }~.
\end{equation}
For the last term in the parenthesis of Eq. (\ref{psiuuexact}), 
analogously to what observed for Eq. (\ref{psiuexact}),    one can use the
transformation given in Eq. (\ref{simpl1}) of Appendix \ref{reduct1}.
In this way,
a momentum dependent term and the spin-orbit interaction are obtained.

\noindent
A case of special interest is when
\begin{equation}\label{vm0}
V_L(\vec r)=0~.
\end{equation}
In this case, 
due to the definition of Eq. (\ref{vpm}), one has $V^0_v(\vec r)=V_s(\vec r)$.
Furthermore, Eq. (\ref{psilfpsiu}) does not depend on $V_L(\vec r)$
and, in consequence, we also have:
\begin{equation}\label{ge0}
B(E;\vec r)= F(E;\vec r)=0
\end{equation}
This case, traditionally denoted as \textit{spin-symmetry} case
\cite{spinsym1,spinsym2},
allows for simple solutions of the Dirac equation, 
in which the spin-orbit interaction is absent and
the orbital angular momentum and the spin are 
\textit{decoupled}.\\
We now consider the case in which the absolute values 
of the matrix elements of the adimensional quantity $F(E;\vec r)$
are small.
In this case $B(E;\vec r)$, defined in Eq. (\ref{ger}), 
can be expanded in a power series of $F(E;r)$:
\begin{equation}\label{geseries}
B(E;\vec r)=
\sum_{k=1}^\infty  
[F(E;\vec r)]^{k}~.
\end{equation}
At the first order $(k=1)$ one simply  has:
\begin{equation}\label{geappr}
B(E;\vec r)\simeq F(E;\vec r)~.
\end{equation}
One can replace this relation in Eq. (\ref{psiuuexact})
obtaining an  approximated equation for $|\Phi_U>$.

%
\section{The one-body reduction}\label{redpro}
Let us now study the formal reduction procedure suitable for the
generalization  to the two- and n-body cases.\\
In the first place, we write the four-component Dirac spinor, that represents 
the \textit{correlated} (approximate) solution, in the form:
\begin{equation}\label{psiapprox}
|\Psi_{corr}>= N_{(1)} \cdot K \cdot |\Phi>
\end{equation}
where $N_{(1)}$ represents the one-body numerical normalization constant, 
to be discussed in the following, $K$ is the  local reduction  operator
that transforms the (reduced) spinor $|\Phi>$ into a four-component Dirac spinor;
it is defined as:

\begin{equation}\label{defk}
K= K(m,E; \vec p, \vec \sigma)=
\begin{pmatrix} 1 \\ 
                {\frac {\vec \sigma \cdot \vec p} {m+E}  }
\end{pmatrix}~.
\end{equation}
In Eq. (\ref{psiapprox}), we have taken as  approximate solution  
a Dirac spinor that represents the exact solution 
in the  case of Eq. (\ref{vm0}), that is when $V_L(\vec r)=0$.	\\
The operator $K$ of the last equation defines the \textit{correlation}
between the upper and lower components of $|\Psi_{corr}>$ .\\
We now replace 
in the original Dirac equation (\ref{dirac1}) (written by means of the Dirac operator
  $D$ of Eq. (\ref{dirop})) 
the exact solution
$|\Psi>$ with
$|\Psi_{corr}>$;
furthermore, in order to obtain
an Hermitean reduced operator acting on $|\Phi>$,
we also multiply the same equation from the left
by $K^\dag $.\\
In consequence, the reduced (approximated) equation for $|\Phi>$ is formally written  in the form:
\begin{equation}\label{appreqphi}
K^\dag  \left[D +W_{(1)} \right]
K ~|\Phi>=0~.
\end{equation}
For the Dirac operator $D$ of Eq. (\ref{dirop}), with standard calculations,
one finds the corresponding  reduced \textit{noninteracting} operator $\hat D$, 
in the form:
\begin{equation}\label{khmek}
\hat D=\hat D (m,E;\vec p )=
K^\dag D K=
m -E  +
{\frac {{\vec p}^2} {m+E} } ~.
\end{equation}
The one-body reduced interaction is written, in general, in the form:
\begin{equation}\label{kwk}
K^\dag W_{(1)} K =\hat W_{(1)}~.
\end{equation}
For the specific Dirac interaction  of Eq. (\ref{intgg1}),
the one-body reduced interaction takes the form:											
\begin{equation}\label{kwkspec}
\begin{split}
\hat W_{(1)}
=V_U(\vec r)+ 
{\frac {1} {m+E}}
\vec \sigma \cdot \vec p       
F(E;\vec r)
\vec \sigma \cdot \vec p= \\
=V_U(\vec r)+ 
{\frac {1} {(m+E)^2 }}
\vec \sigma \cdot \vec p       
V_L(\vec r)
\vec \sigma \cdot \vec p~~.
\end{split}	
\end{equation}
The reduction procedure can be generalized to  any interaction.
The whole Appendix \ref{reduct1} is devoted to calculate the  reduction
of the one-particle interaction with  external scalar and vector fields. 
At the end of that Appendix, 
the corresponding transformation equations are also given.\\
Note that  $\hat W_1$, in the previous equation, and also, in the following,
the two- and n-body reduced interactions $\hat W_{(2)}$ , $\hat W_{(n)}$
are all \textit{energy dependent} operators.

\noindent
Considering  Eqs. (\ref{khmek}) and (\ref{kwk}),
we can write the one-body Dirac reduced equation in the form:
\begin{equation}\label{dirred1}
\left [ \hat D + \hat W_{(1)} \right] |\Phi>=0~.
\end{equation}
We have obtained for $|\Phi>$ the same equation
derived for $|\Phi_U>$, see Eq. (\ref{psiuuexact}),
 with $B(E;\vec r)$  expanded up to the order  $k=1$, as given in Eq. (\ref{geappr}).

\vskip 0.5 truecm
\noindent
We now introduce:
\begin{equation}\label{ksquared}
\hat Q= \hat Q(m,E;\vec p)=
K^\dag  K =
1 + 
{\frac  {{\vec p}^2}   {(m+E)^2} }~.
\end{equation}
By means of this operator,
we can  define
the one-body normalization constant, $N_{(1)}$, 
that is  unrelevant for 
obtaining the energy eigenvalue $E$
but is necessary to determine in a complete way the correlated Dirac spinor
and  to calculate the matrix elements of any (other) Dirac operator.
The normalization constant $N_{(1)}$  can be obtained by requiring that,
for a bound state, the correlated Dirac spinor of Eq. (\ref{psiapprox})
is normalized to unity.
By using  Eq. (\ref{ksquared}),
one has the following implicit definition:
\begin{equation}\label{norm1}
1=  N_{(1)}^2 <\Phi| \hat Q |\Phi>=
    N_{(1)}^2   \int d^3 r ~\Phi^\dag(\vec r)~ \hat Q ~\Phi(\vec r)
\end{equation}
 from which one can immediately obtain $N_{(1)}$.
We consider $  N_{(1)}$ as a numerical constant, not included in the 
definition of $K$, in order to have a \textit{local} reduced Dirac equation.
Otherwise, one could introduce the normalized, nonlocal, reduction operator, 
in the form:
\begin{equation}\label{knorm}
K_{norm}=K \cdot \left [ 1+ {\frac  {{\vec p}^2}   {(m+E)^2} } \right] ^{-1/2}~.
\end{equation}
This choice will not be used in the present work because we prefer
to obtain a local equation.\\
Finally, we anticipate that 
$N_{(2)}$ and $N_{(n)}$ that respectively represent
the two-body and the n-body normalization constants,
will be determined with an analogous procedure
in Sect. \ref{twobody}.  
 
\noindent
We also note that the \textit{exact} equation  (\ref{psiuuexact}),
without expansion of $B(E;\vec r)$, can be
recovered if in the interaction
(see Eq. (\ref{kwkspec}) )
one replaces $V_L(\vec r)$  with $V_L^{eff}(\vec r) $ defined as:
\begin{equation}
V_L^ {eff}(\vec r)  = V_L(\vec r)\cdot 
{\frac {1} {1- {F(E;\vec r)} } }~.
\end{equation}
Otherwise, if the fundamental interaction is not known, 
one can construct a phenomenological model  for $W_{(1)}$,
by using  a suitable parametrization  and then fitting the results 
to the experimental data.\\
 
\section{Two- and n-body reduction of the DLE}\label{twobody}
We introduce here the generalization of our model to the two- and n-body case.
We start analyzing in detail the (relatively simple) two-body case.
The DLE  is formally written in form:
\begin{equation}\label{dirac2}
[D_1 + D_2  +W_{(2)}]|\Psi>=0
\end{equation}
where we have used, for each particle ($i=1,2$), the standard one-body  Dirac
operator 
defined in Eq. (\ref{dirop})  with the shorthand notation introduced in
Subsect. \ref{symbnot}.
Furthermore, 
$W_{(2)}$ represents the Dirac interaction operator for the two-body case
and $|\Psi>$ is the  Dirac state of the system.
Finally,
the total energy is 
$E_T=E_1+E_2$. 
%
%
\vskip 0.5 truecm
\noindent
In a relativistic context, the separation of variables into CM and relative
variables is a difficult problem
that will not be studied  here in detail.
In order to calculate the mass $M$ of the two-body bound system, 
it is sufficient to
study the problem in the Center of Mass (CM) reference frame, where $E_T=M$
and the total momentum $\vec P$ is vanishing.
In this respect, 
\textit{without introducing a new notation}, we assume in the following that
 all the states we use 
(\textit{i.e.}, Dirac states, correlated Dirac states and reduced states)  
satisfy the condition of vanishing momentum: 
\begin{equation}\label{condtotzero}
\vec P | \Psi>=0,~\vec P | \Psi_{corr}>=0,~ \vec P | \Phi>=0~ .
\end{equation}
In order to define the relative variables 
%
we shall  focus our attention on a specific, relatively simple, case
that   corresponds directly to the very relevant physical
systems of the $ q \bar q$ mesons.
(However, as we shall see in the following, the formal reduction procedure of our model
is quite general and does not depend on the specific choice of the CM and relative variables.)\\
Now we consider two equal mass particles:
\begin{equation}\label{eqmass}
m_1=m_2=m~,
\end{equation}
furthermore,  we assume that, in the CM, the two particles have the same energy:
\begin{equation}\label{eqen}
E_1=E_2= {\frac {E_T} {2}}= {\frac {M} {2}}  ~.
\end{equation}
The momentum operators of the two particles are given by:
\begin{equation}\label{momentumops}
\vec p_1= -\vec p, ~~ \vec p_2= \vec p
\end{equation}
where $\vec p$ represents the relative momentum operator (in the CM reference frame), canonically conjugated
to the relative distance vector 
\begin{equation}\label{rdv}
\vec r =\vec r_2 - \vec r_1~.
\end{equation}
In this way,
we can introduce, in that frame,
 the Dirac wave function
 $\Psi(\vec r)=<\vec r | \Psi> $;
furthermore, in a local model,
the interaction operator  depends on the
spatial variable $\vec r$, that is $W_{(2)}=W_{(2)}(\vec r)$.\\
We construct the reduced equation by introducing, for the Dirac correlated wave function, the following expression:

\begin{equation}\label{psiapprox2}
|\Psi_{corr}>= N_{(2)} \cdot K_1 \cdot K_2
\cdot | \Phi>
\end{equation}
where  
$K_i$ 
represents the  one-particle reduction operator of the $i$-th particle  ($i=1,2$), 
as given
in Eq. (\ref{defk}).
Specifically, for the \textit{arguments} of these operators 
(and of all the other operators in the following)
 the definitions  of Eqs. (\ref{eqmass})-(\ref{momentumops}) are used.
Finally, in Eq. (\ref{psiapprox2}),
$|\Phi>$ is the 
two-particle reduced  state.
Finally $N_{(2)}$ is the numerical 
two-body normalization constant.
This last quantity is implicitly defined by 
normalizing  $|\Psi_{corr}>$ to unity, that is:
\begin{equation}\label{n2impdef}
1= N_{(2)}^2 <\Phi| \hat Q_1 \hat Q_2 |\Phi>=
N_{(2)}^2
\int d^3 r \Phi^\dag(\vec r) \hat Q_1 \hat Q_2
\Phi(\vec r)~.
\end{equation}
By using Eqs. (\ref{eqmass})-(\ref{momentumops}),
 for a two-body, equal mass system,
one has $\hat Q_1=\hat Q_2$ and
\begin{equation}\label{q1q2presc}
\hat Q_1 \cdot \hat Q_2= 
 \left[ 1 + {\frac {\vec p^2} {(E_T/2+m)^2} } \right]^2~.
\end{equation}
As in the one-body case,
after replacing $|\Psi_{corr}>$ in Eq. (\ref{dirac2}),
 we multiply from the left the same equation
by $  K_1^\dag \cdot K_2^\dag $
in order to obtain an Hermitean reduced operator.
We have:
\begin{equation}\label{dir2red1}
 K_1^\dag \cdot K_2^\dag
(D_1 + D_2  +W_{(2)})
       K_1
 \cdot K_2
| \Phi> =0~.
\end{equation}
Using for the one-body operators $\hat Q_i$ and $\hat D_i$
their definitions of Eqs. (\ref{ksquared}) and (\ref{khmek}) respectively, 
the previous equation can be rewritten as:
\begin{equation}\label{dir2red2}
\left[ \hat Q_2 \hat D_1 + \hat Q_1 \hat D_2 +\hat W_{(2)} \right]
| \Phi> =0
\end{equation}
where the two-body reduced interaction is:

\begin{equation}\label{w2red}
\hat W_{(2)}= K_1^\dag \cdot K_2^\dag ~ W_{(2)}
~ K_1 \cdot K_2~.
\end{equation}
The reduction of a scalar and vector two-body interaction
is studied in detail in Appendix \ref{reduct2}.

%
%
%
\noindent
With the specific definitions for the arguments of the operators,
given in Eqs.  (\ref{eqmass})-(\ref{momentumops}),
one has $\hat Q_1= \hat Q_2 $ and $\hat D_1= \hat D_2$;
in consequence,
the explicit reduction of the noninteracting operator, in the CM, gives: 
\begin{equation}\label{kterm2}
\begin{split}
\hat Q_2 \hat D_1 + \hat Q_1 \hat D_2=
-\hat  G_{(2) D}^{-1}(E_T)=~~~~~~~~\\
=\left[ 1 + {\frac {\vec p^2} {(E_T/2+m)^2} } \right] 
\left( {\frac { 2 \vec p^2} {E_T/2+m} } +2m -E_T \right)
\end{split}
\end{equation}
where, analogously Eq. (\ref{gm2dl}), we have introduced 
the shorthand notation $\hat  G_{(2) D}^{-1}(E_T)$
for the reduced operator, inverse of the Green function.
In this way, the reduced equation can be written as:
\begin{equation}\label{eq2fdl}
\left[ -\hat  G^{-1}_{D}(E_T) + \hat W_{(2)}
\right ] |\Phi>=0
\end{equation}
The explicit expression of Eq. (\ref{kterm2}) clearly shows that our model 
does not admit any free solution with $E_T=0$, avoiding the CDP.
Also note that both the reduced interaction of Eq. (\ref{w2red}) and
the operator of Eq. (\ref{kterm2}) are \textit{local} quantities.

\vskip 0.5 truecm
\noindent
From the previuos discussion, one can  easily find  the generalization 
to the case of a system with $n$ constituents.
The DLE has the form:
\begin{equation}\label{diracn}
\left [ \sum_{i=1}^n D_i + W_{(n)} \right ] |\Psi>=0
\end{equation}\\
In the CM frame, one has to introduce
as spatial variables  
the set of $n-1$ Jacobi variables, collectively denoted as $ \{\vec r\}$
and their conjugated Jacobi momenta  $ \{\vec p\}$.
Furthermore, one has to express  
the particle momenta
$\vec p_i$ in terms of the Jacobi momenta.
All the states satisfy, in the CM, the vanishing momentum condition 
(\ref{condtotzero}).
The Dirac correlated state is defined as:
\begin{equation}\label{dircorrn}
|\Psi_{corr}>= \prod_{j=1}^n K_j
 \cdot |\Phi> ~.
\end{equation}
Then, the reduction of the DLE is performed analogously
 to Eq. (\ref{dir2red1}), giving:

\begin{equation}\label{dirnred1}
 \prod_{i=1}^n K_i^\dag \cdot
\left [ \sum_{k=1}^n
D_k + W_{(n)}
\right ]
 \cdot 
\prod_{j=1}^n K_j
 \cdot |\Phi> =0~.
\end{equation}
In consequence,
the reduced equation, that generalizes Eq. (\ref{dir2red2}), takes the form:
\begin{equation}\label{dirnred2}
\left[ 
\sum_{i=1}^n
\left(\prod_{j\neq i}^n
\hat Q_j
\right)
 \hat D_i  +\hat W_{(n)} \right]
 |\Phi> =0
\end{equation}
where the product is performed over all the $n$ particles, excluding
the $i$-th one. 
The reduced interaction is:
\begin{equation}\label{weffndef}
\hat W_{(n)} =
\prod_{i=1}^n K_i^\dag \cdot W_{(n)} \cdot
\prod_{j=1}^n K_j
\end{equation}
Finally, the  implicit normalization condition for the reduced wave function is:
\begin{equation}\label{nnimpdef}
1= N_{(n)}^2 
<\Phi|\prod_{i=1}^n \hat Q_i |\Phi>=
N_{(n)}^2 
\int d^3 \{r\}
 \Phi^\dag(\{\vec r\} )~ \prod_{i=1}^n \hat Q_i
~\Phi(\{\vec r \} )
\end{equation}
that generalizes the two-body case of Eq. (\ref{n2impdef}).\\
Obviously, all the expressions for the n-body reduced operators become increasingly 
more complex as $n$ increases.
%
\section{Reduction of the MWE}\label{applmw}
The present reduction procedure can be also applied to the two-body MWE.
We recall that this equation avoids from the beginning the CDP by including 
in the definition of the Green function the so-called \textit{crossed graphs},
as discussed in Appendix \ref{tr2bwe}.
For the two-body case, the MWE takes, in our notation, the following form:
\begin{equation}\label{mw2}
\left [D_1 S_2 + D_2 S_1  +W_{(2)} \right]|\Psi>=0
\end{equation}
where the first two terms represent
the noninteracting operator given by $-G_{(2)M}^{-1}$ of Eq. (\ref{gm2mw}).
The MWE 
should be compared with the DLE of Eq. (\ref{dirac2}),  
analyzed  in Appendix \ref{tr2bwe}. 
In particular, the difference with respect to that equation consists in the
insertion of the energy-sign operators $S_i$
(denoted as $\hat \rho_i$ in the original paper \cite{mwa}).
These operators are introduced in  Eq. (\ref{ensign})
and are calculated here for the $i$-th particle.
When applied to the free Dirac spinors, they give
the energy sign of the free particle, as shown in Eq. (\ref{ensignlamb}).
Note that, due to the presence of the $S_i$,
in the MWE it is not possible to introduce an Hamiltonian operator.\\
%
%
\noindent 
The reduction of the MWE (\ref{mw2}) is performed with the same technique
used for the DLE in Sect. \ref{twobody} .
The vanishing momentum condition (\ref{condtotzero}) is used.
Also, the same definitions of Eqs. (\ref{eqmass})-(\ref{rdv}) for the two-body equal mass problem are used here.
The correlated Dirac state is given by Eq. (\ref{psiapprox2}).
Analogusly to Eq. (\ref{dir2red1}), we have:

\begin{equation}\label{mw2red1}
 K_1^\dag \cdot K_2^\dag
\left [ D_1 S_2 + D_2 S_1  +W_{(2)} \right]
       K_1
 \cdot K_2
 |\Phi> =0~.
\end{equation}
This reduced equation can be rewritten as:
\begin{equation}\label{mw2red2}
\left[\hat D_1 \hat S_2  +\hat D_2  \hat S_1  +\hat W_{(2)} \right]
 | \Phi> =0
\end{equation}
that replaces the Dirac-like reduced equation (\ref{dir2red2}).
In the previous equation we have introduced the reduced  $\hat S_i$ operators.
The reduced $\hat S$  operator has the  general form:
\begin{equation}\label{hatr}
\begin{split}
\hat S= \hat S(m, E;\vec p)=
K^\dag S K= ~~~~~~~~~~~~~~\\
={\frac 1 {\varepsilon}}
\left [ \hat D + E \hat Q \right ] =
{\frac 1 {\varepsilon }}
\left [ m + {\frac {\vec p^2} {m+ E}} + {\frac { \vec p^2 E} {(m+ E)^2}}
\right ] ~.
\end{split} 
\end{equation}
With the definitions of  Eqs. (\ref{eqmass})-(\ref{momentumops})
 one has $\hat S_1= \hat S_2$ and $\hat D_1= \hat D_2$.
The reduction of the noninteracting  operator 
 that appears in Eq. (\ref{mw2red2}),
 takes the form:
\begin{equation}\label{kterm2mw}
 \begin{split}
\hat D_1 \hat S_2  +\hat D_2  \hat S_1 = -\hat G^{-1}_{(2) M}(E_T)= ~~~~~~~~~~~~~~\\ 
={\frac 1 \varepsilon}
\left [ m + {\frac {\vec p^2} {m+ E_T/2}} + {\frac { \vec p^2  E_T/2 }{(m+ E_T/2)^2}}
\right ]
\left( {\frac { 2 \vec p^2} {E_T/2+m} } +2m -E_T \right)
\end{split}
\end{equation}
where, in analogy with Eq. (\ref{kterm2}), 
we have  introduced the shorthand notation $\hat G^{-1}_{(2) M}(E_T)$.
Inserting Eq. (\ref{kterm2mw}) in Eq. (\ref{mw2red2}) one can write the complete
reduced equation as:
\begin{equation}\label{eq2fmw}
\left[ -\hat G^{-1}_{(2) M}(E_T) + \hat W_{(2)}
\right ] |\Phi>=0~.
\end{equation}
Finally, the reduced interaction $\hat W_{(2)}$ is the same as that of the DLE, 
given in Eq. (\ref{w2red});
the same
implicit normalization condition of Eq. (\ref{n2impdef})
is used here.\\

\section{Model of  $q ~ \bar q$ interaction  for the Charmonium spectrum}\label{charm}
We shall apply our reduction to the study of Charmonium spectrum.
In the present work we consider only a relatively simple effective interaction
model, following the standard prescriptions used for the study of heavy quarkonia.
A study of different possible interactions with an analysis
of the  physical meaning of the various terms must be performed in a work apart.

\noindent
In the present model,
for the two-body interaction  $W_{(2)}$, 
we take   the sum  of a vector and scalar term, that is:
\begin{equation}\label{wmodel}
W_{(2)}= W_{(2)}^v+ W_{(2)}^s~.
\end{equation}

\vskip 0.5 truecm
\noindent
For the vector interaction we take the following standard expression:
\begin{equation}\label{vvect}
W_{(2)}^v=V_{(2)}^v(r)
\gamma_1^0 \gamma_2^0 \cdot
\gamma_1^\mu \gamma_2^\nu g_{\mu \nu}
\end{equation}
with the Dirac matrices recalled in Subsect. \ref{symbnot}.
\noindent
The potential function $V_{(2)}^v(r)$ will be discussed in the following.\\
In order to have a local interaction operator we have not included the retardation 
contributions.
This approximated choice can be considered consistent with
Eq. (\ref{eqen}):
we make the hypothesis that the quark energies are 
 \textit{fixed}; 
consequently, the quarks do not interchange energy 
with the effective gluonic field that mediates the interaction.  

\noindent
For the scalar interaction we take the expression:
\begin{equation}\label{vscal}
W^s_{(2)}=V_{(2)}^s(r)
\gamma_1^0 \gamma_2^0 ~.
\end{equation}
We now discuss the spatial potential functions 
$V_{(2)}^v(r)$ and $V_{(2)}^s(r)$  of the model.
For   the vector potential function of Eq. (\ref{vvect}),
we take the following effective, regularized, expression: 
%
\begin{equation}\label{vpot}
V_{(2)}^v(r)=\bar V_v ~ -{\frac 4 3} \cdot {\frac {\alpha_v} {r}} \cdot F_v(r)
\end{equation}
where $4/3$ is the color factor and $\alpha_v \equiv \alpha_{strong}$ represents the 
effective strong coupling constant; we use the subscript $v$ (that denotes the vector interaction)  to avoid confusion with the scalar terms.\\
The regularization for $r\rightarrow 0$ is performed having in mind a non-pointlike 
chromo-electric charge distribution of the quarks
that gives rise to the additive energy constant 
$\bar V_v$ and to the regularization function  $ F_v(r)$, 
as shown in Ref. \cite{chromomds}.
A detailed study of the relationship between these two quantities must be done
in a different  work.
Here we recall that $\bar V_v$ is introduced also to reproduce
phenomenologically the quark confinement. 
As for $F_v(r)$, 
we choose
\begin{equation}\label{fvreg}
F_v(r)=\text{erf} \left({\frac r d_v} \right)
\end{equation}
being $d_v$ the regularization range.
Note that $F_v(\infty)=1$, not altering the long distance Coulombic behaviour,
 and, for $r \rightarrow 0$, $F_v(r)\simeq {\frac {2} {\sqrt{\pi}}}{\frac r  d_v} $;
in this way the Coulombic singularity  is eliminated.

\vskip 0.5 truecm
\noindent
As for the scalar interaction, after trying different expressions, we take 
the potential function of Eq. (\ref{vscal}) in 
the following form:
\begin{equation}\label{spot}
V_{(2)}^s(r)=\bar V_s {\frac 1 2} 
\left[
\text{erf} \left( (r-r_s)/d_s \right) -1
\right]~.
\end{equation}
Note that this potential represents a hole of depth approximately equal to
$ - \bar V_s$ at $r=0$, while for $r \rightarrow \infty$, one has $V_s=0 $;
%
the width of the hole is approximately  $ r_s$;
 finally, the parameter $d_s$ is related to the squareness of the hole. 
\vskip 0.5 truecm
\noindent
As it will explained in Sect. \ref{charmres}, we shall use different numerical values 
for some parameters of the model in order to reproduce accurately the
resonances above the open charm threshold.\\
%
The reduced interactions $\hat W_{(2)}^v$ and $\hat W_{(2)}^s$ for our equation
are obtained reducing the expressions of
Eqs. (\ref{vvect}) and (\ref{vscal}), respectively.
To this aim, we use 
the two-body reduction equations of Appendix \ref{reduct2}, 
specifically:
Eq. (\ref{scalvtred2b}) for the product of the time components
  ($t$) of the vector interaction
and for the scalar ($s$) interaction;
Eq. (\ref{vectred2b}) for the product of the spatial parts of the vector interaction.
 
\section{Solution method}\label{solution}
In order to solve Eq. (\ref{eq2fdl}) for the DLE and 
Eq. (\ref{eq2fmw}) for the MWE we use a variational procedure,
introduced in Ref. \cite{mdsquint},
that consists in diagonalizing the operators of the equations in  
a HO basis.
The trial wave functions of this basis  can be written
in the coordinate space as:
\begin{equation}\label{basisho}
\Phi_{n; L,S,J}(\vec r)=< \vec r|  n; L, S, J> =
R_{n,L}(r;\bar r) {[Y_L(\hat r) \otimes \chi_S]}_J ~.
\end{equation}
In the previous equation the trial radial function is represented by
$R_{n,L}(r;\bar r)$, being $n$ the principal  HO quantum number and
$\bar r$ the variational parameter with the dimension of longitude;
 $Y_{L,M_L}(\hat r)$ is the corresponding spherical harmonic
and $\chi_{S,M_S}$, with $S=0,1$ is the  $c~ \bar c$
coupled spin function.
The orbital angular momentum and the spin are standardly coupled to 
the total angular momentum $J,M_J$.
For brevity we do not write $M_J$ because it is unrelevant for the calculations
of rotationally scalar operators.\\
Furthermore,
for simplicity reasons, we do not consider the possibility of mixing between 
states with different values of $L$ , because these effects have been shown to be negligible in semirelativistic calculations.\\
The radial HO functions have the explicit form:
\begin{equation}\label{radialho}
 R_{n,L}(r;\bar r) = {\frac {1} {{\bar r}^{\frac 3 2}}   }
\left [{\frac {2(n !)} {\Gamma(n+L+{\frac 3 2} )} }
\right ]^{\frac 1 2}
s^L {\cal L}_n^{L+ {\frac 1 2}}(s^2)
\exp \left(- {\frac {s^2} {2}} \right)
\end{equation}
where $s=r/\bar r$ is the adimensional variable and 
${\cal L}_n^{L+ {\frac 1 2}}(s^2)$
are the generalized Laguerre polynomials.\\

\noindent
The matrix elements  of the operator  $\hat G^{-1}_{(2) D}(E_T)$ 
of Eq. (\ref{kterm2}), 
can be calculated in the coordinate space, because in that operator there only appear
 finite powers of the momentum operator, 
that is
 $\vec p^{2q}$, with $0 \leq q\leq2$.
On the contrary, $\hat  G^{-1}_{(2) M}(E_T)$ of Eq. (\ref{kterm2mw}),
due to the factor $1/\varepsilon$, depends \textit{nonlocally} on the momentum;
in consequence its matrix elements must be evaluated in the momentum space.
To this aim we use the standard analytic expression of  the
HO wave functions in the momentum space
$\Phi_{n; L,S,J}(\vec p)=< \vec p|  n; L, S, J> $.
We recall that in both DLE and MWE, $\hat W_{(2)} $ is a local operator
whose matrix elements are calculated in the coordinate space.
In particular,
the $\vec \sigma_i \cdot \vec p_i$ operators are applied to the  wave functions of Eq. (\ref{basisho}).
As explained in Appendix \ref{tr2bwe},
this procedure would not be possible if positive (and negative) energy 
projectors where used, requiring, in any case,  
an integral equation in the momentum space.\\
We note that our reduced equations (\ref{eq2fdl}) and (\ref{eq2fmw})
do not represent standard eigenvalue equations.
On the contrary, due to the reduction procedure,
 $\hat G^{-1}_{(2) D}(E_T)$ and $\hat  G^{-1}_{(2) M}(E_T)$
(given in Eq. (\ref{kterm2}) and Eq. (\ref{kterm2mw}), respectively),
and also $\hat W_{(2)}$,
\textit{depend} on the total energy $E_T$; 
consequently, we have to solve for both models
 an energy dependent equation.\\
To this aim we make the following replacement for
$\hat  G_{(2) X}^{-1}(E_T)$ : 
\begin{equation}\label{gm1e}
-\hat G^{-1}_{(2)X} (E_T)= \hat {\cal F}_X(E_T) -E_T
\end{equation}
where the subscript $X=$ $D,~M$ stands for DLE or MWE. \\
In this way the energy dependent equation can be formally written as:
\begin{equation}\label{eqfet}
\left [
\hat {\cal F}_X(E_T) +\hat W_{(2)}(E_T)
\right ]
|\Phi> =E_T | \Phi>~.
\end{equation}
We replace $E_T$ in the \textit{l.h.s.} with the auxiliary parameter $E_V$,
obtaining the following fictitious eigenvalue equation: 
\begin{equation}\label{eqfetfict}
\left [
\hat {\cal F}_X(E_V) +\hat W_{(2)}(E_V)
\right ]
|\Phi> =E_T |\Phi>~.
\end{equation}
We can solve \textit{variationally} this equation (as explained below)
 for a given $E_V$ and determine the 
corresponding value of  $E_T$ in the $r.h.s.$.
Then, we vary $E_V$ until the value found for  $E_T$ is equal to $E_V$ of the $l.h.s.$.
This value gives the solution of Eq. (\ref{eqfet})
and represents  the energy  of the system. \\
As for the variational procedure to solve the fictitious  eigenvalue equation
(\ref{eqfetfict}),
we obtain good numerical convergence for $E_T$,
taking the first ten trial wave functions of the basis for each state.
In more detail, the $10 \times 10$ \textit{l.h.s} matrix  is diagonalized and minimized
by means of the standard variational approach \cite{mdsquint}.

\section{Study of the Charmonium spectrum }\label{charmres}
In this Section we apply the reduced DLE and MWE  to study the Charmonium spectrum
with the interaction introduced in Sect. \ref{charm}.
The obtained theoretical results and the experimental data \cite{pdg} are shown 
in Table \ref{tabres}; 
the values of the parameters used for the calculation 
are given in Table \ref{tabpar}.\\ 
The present model, that takes into account a \textit{fixed} number of degrees of freedom, is expected to work properly for the resonances below the open charm threshold.
For higher resonances some mechanism that takes into account the creation of new particles should be implemented.\\
We consider here the very simple, purely phenomenological, strategy  of taking 
different values of some parameters of the interaction 
above the open charm threshold.
In more detail,
we introduce three intervals for the values of the resonance mass $M$ of the spectrum.
These intervals,
 $I_1$, $I_2$ and $I_3$, are defined as:\\
- $I_1, ~~~~~ M < M_a$ , \\
- $I_2, ~~~~~ M_a \leq M < M_b$ ,\\
- $I_3, ~~~~~ M \geq M_b $ \\
where  $M_a$ corresponds to  the open charm threshold and $M_b$ has been fixed,
after some trials, to obtain a good reproduction of the data.
Their values are given in Table \ref{tabpar}. 
As shown in Table \ref{tabres},
in the interval $I_1$ we have considered all the eight experimentally observed resonances;
in the intervals $I_2$ and $I_3$ we have considered respectively
five and three
not controversial resonances.
For a discussion about the phenomenological interpretation of the resonances
in different models, the interested reader is referred to Ref. \cite{mdsf}. \\
In principle, the parameters of the model are the quark mass $m_q$ and
the interaction parameters, introduced in Sect. \ref{charm}, that are:
$\alpha_v(I_i), ~ \bar V_v(I_i), ~ d_v(I_i),$
$\bar V_s(I_i), ~ r_s(I_i)$
and $d_s(I_i)$ for the three intervals $I_1,~I_2$ and $I_3$.\\
The quark mass $m_q$
 has not been  considered as a free parameter
but has been fixed at the \textit{current mass} QCD value \cite{pdg},
as shown in Table \ref{tabpar}.\\
We have performed \textit{two fits}, denoted as ``A" and ``B",
with the objective of obtaining an accurate 
theoretical reproduction of whole experimental spectrum with the smallest possible number of free parameters.
To this aim  we have \textit{vinculated} the numerical values of some parameters 
in the different intervals.\\
In more detail, as shown in Table \ref{tabpar},
 in the interval $I_1$, all the interaction parameters
are free parameters of the fit.\\
In the interval $I_2$, $\alpha_v(I_2)$, $\bar V_v(I_2)$ are free parameters;
the vinculated parameters are: 
$d_v(I_2)=d_v(I_1), \bar V_s(I_2)=\bar V_s(I_1), d_s(I_2)=d_s(I_1)$;
in the fit A, $r_s(I_2)$ is a free parameters, while in the fit B,
it is vinculated: $r_s(I_2)= r_s(I_1)$.\\ 
In the interval $I_3$ all the parameters (in both fits A and B) are vinculated 
as follows:
$\alpha_v(I_3)=\alpha_v(I_2)$, $\bar V_v(I_3)=\bar V_v(I_1)$,
$d_v(I_3)= d_v(I_1)$, $\bar V_s(I_3)=\bar V_s(I_1)$,
$r_s(I_3)=r_s(I_1)$ and  $d_s(I_3)=d_s(I_1)$. \\
As a result of the fit procedure,
the \textit{same theoretical masses} have been obtained
 by using  the reduced DLE and MWE; these values, for the two fits, are
 shown in the columns Theor.(A) and Theor.(B) of Table \ref{tabres}.
The results of the fit A, with one more free parameter, are slightly better
than the results of fit B.\\
The values of the free parameters of the fits present small differences
for the two equations, as reported in the columns DLE(A), DLE(B) and MWE(A), MWE(B)
 of Table \ref{tabpar}. \\
Both the reduced DLE and MWE allow for an accurate reproduction of the spectrum,
showing that, for the Charmonium case, there is no argument 
to prefer one of the two equations.
Some more comments are given in the Conclusions.
\section{Conclusions}\label{conclusions}
A local, energy dependent reduction of the DLE has been derived.
The same technique has been also applied to  the MWE, obtaining in both cases
a relativistic equation that can be solved with standard numerical techniques.
Further investigation is needed to relate more strictly the reduced equation
 to the dynamics of the underlying
field theory.\\
The reduced equations have been applied to the study of the Charmonium spectrum
obtaining accurate results.
Both the DLE and the MWE give the same spectrum with small differences of the free parameters.
This result can be related to the reduction procedure: the contributions
of the $+-$, $-+$ and $--$ states, that are \textit{different} for the DLE and MWE,
are diminished by the reduction operators $K_i$ while, in both equations,
 the more relevant contributions
are given by the $++$ states that are the same for the two equations.\\
A deeper study of the interaction, possibly considering different Lorentz structures
beyond the standard vector-scalar model, should be also undertaken.
\vskip 0.5 truecm
\centerline{{\bf Aknowledgements}}
The author thanks the group of  ``Gesti\'on de Recursos de Computo Cient\'ifico,
Laboratorio de Biolog\'ia Computacional,
Facultad de Ciencias - Universidad Nacional de Colombia"
for the access to the cluster that was used to perform the numerical calculations 
of this work.

\vskip 0.5 truecm
\appendix
\section{Three-dimensional   two-body wave equations}\label{tr2bwe}
We give here some technical details about the
TDRWEs for two-body bound systems related to the present work.\\
\textit{One-body case}. 
We start from some relevant  one-body quantities.\\
The spinors for a free Dirac particle of momentum $\vec p$
(omitting the two component spin factor), 
can be written the form:
\begin{equation}\label{upm}
u_\lambda=
u_\lambda(m; \vec p, \vec \sigma) =M \cdot U_{\lambda} 
\end{equation}
where $\lambda=\pm 1$ is the energy sign;
the $ U_{\lambda} $
are written as:
\begin{equation}\label{defkpm}
U_+ = U_+(m; \vec p, \vec \sigma)=
\begin{pmatrix} 1 \\ 
                {+\frac {\vec \sigma \cdot \vec p} {\varepsilon+m}  }
\end{pmatrix} 
~~~~U_- = U_-(m; \vec p, \vec \sigma)=
\begin{pmatrix}
{-\frac {\vec \sigma \cdot \vec p} {\varepsilon+m}  }\\
1 
\end{pmatrix}
\end{equation}
where \begin{equation}\label{eps} 
\varepsilon=\varepsilon(m;\vec p) =+\sqrt{{\vec p}^2 + m^2}
\end{equation} 
is the on-shell
positive energy of the particle. The factor
\begin{equation}\label{mdef}
M= M(m;\vec p)=\sqrt{ {\frac {\varepsilon+m} {2 \varepsilon} }}
\end{equation}
normalizes to 1 the spinors, 
that is $u_\lambda^\dag u_\xi=\delta_{\lambda \xi}$.
Note that the $ U_{\lambda} $, due to  $\varepsilon$, 
depend \textit{nonlocally} on the momentum $\vec p$.\\
The spinors of Eq. (\ref{upm}) obviously diagonalize the free Dirac Hamiltonian:
\begin{equation}\label{diagpm}
{u ^\dag}_\lambda H^{free}~ u_\xi=
\lambda \cdot \delta_{\lambda \xi}\cdot\varepsilon~.
\end{equation}
We also introduce here the energy-sign operator
\begin{equation}\label{ensign}
S=  S(m;\vec p, \vec \alpha, \beta)=
{\frac 1 \varepsilon} \cdot H^{free}
\end{equation}
that, applied to the free spinors, gives:

\begin{equation}\label{ensignlamb}
S u_\lambda = \lambda u_\lambda~.
\end{equation}
The  operator $S$ appears in the MWE and
 will be used in the  following
when discussing that three-dimensional relativistic equation.\\
We introduce the one-particle  projection operators onto positive ($\lambda=+1$)
and  negative ($\lambda=-1$)  energy states:
\begin{equation}\label{prostpm}
\Lambda^\lambda= \Lambda^\lambda(m;\vec p,\vec \alpha, \beta)=
{\frac {1}  {2 \varepsilon}}
  (\varepsilon  +\lambda H^{free})  
	={\frac {1}  {2 }}(1 + \lambda S)
	= \sum_\lambda u_\lambda u_\lambda^\dag   
\end{equation}
where $\varepsilon$ is the relativistic particle energy, defined in Eq. (\ref{eps}),
and $H^{free}$ is the free Dirac Hamiltonian of Eq. (\ref{free1}).\\

\noindent
One can use  the positive and negative energy Dirac spinors
$u_\lambda$ of Eq. (\ref{upm})
to rewrite the original Dirac equation (with interaction)
as a coupled equation for positive and negative energy components.\\
A generic Dirac  state, decomposed into the two spinors of Eq. (\ref{upm}),
 is written, in the \textit{ket} notation, as
\begin{equation}\label{psidirpm}
|\Psi>= \sum_\xi  u_\xi|\Phi_\xi> ~.
\end{equation}	
We recall that the two-component  wave functions, in the momentum space, 
are standardly written as:
$		\Phi_\xi(\vec p)= <\vec p 	|\Phi_\xi>	$.	\\
We consider the Dirac equation (\ref{dirac1}) for the state of Eq. (\ref{psidirpm})
and multiply from the left by
$u_\lambda^\dag$.
We also introduce the \textit{projected interaction operator}
$W_{(1)}^{\lambda, \xi}={u ^\dag}_\lambda W_{(1)}~ u_\xi$.
With standard calculations and
 using Eq. (\ref{diagpm}) one obtains the following coupled   equations:
\begin{equation}\label{diracpmbk}
\sum_\xi
\left [(\lambda \cdot \varepsilon -E)\delta_{\lambda \xi}+ 
W_{(1)}^{\lambda, \xi}
\right] |\Phi_\xi>=0~.
\end{equation}
To solve this coupled equation it is not possible to use the coordinate space.
Even if a \textit{local} interaction $W_{(1)}=W_{(1)}(\vec r)$ were considered,
the presence of the \textit{nonlocal} 
$u ^\dag_\lambda$ and $u_\xi$ 
in $W_{(1)}^{\lambda, \xi}$
requires, in any case, to use  the momentum space.
Defining
\begin{equation}\label{wproj}
{\cal W}_{(1)}^{\lambda, \xi}(\vec p, \vec p')=
<\vec p| {u ^\dag}_\lambda W_{(1)}~ u_\xi |\vec p'>
\end{equation}
one obtains the following coupled integral equations: 
\begin{equation}\label{diracpm}						
\sum_\xi \left [
(\lambda \cdot \varepsilon -E)\delta_{\lambda \xi} 
 \Phi_\xi(\vec p) +
\int d^3 p' 
 {\cal W}_{(1)}^{\lambda, \xi}(\vec p, \vec p')
\Phi_\xi(\vec p') \right ] =0~.
\end{equation}

\noindent
If the off-diagonal matrix elements  can be considered small,
that is
\begin{equation}\label{vanish}
{\cal W}_{(1)}^{\lambda, \xi}(\vec p, \vec p') \simeq 0
~~~~\text{for}
\lambda \neq \xi ~
\end{equation}
and also $\Phi_-(\vec p)$ is negligible,
one obtains an \textit{approximate}  equation for $\Phi_+(\vec p)$ in the form:
\begin{equation}\label{diracpp}						
 (\varepsilon -E) \Phi_+(\vec p) +
\int d^3 p' 
{\cal W}_{(1)}^{+, +}(\vec p, \vec p') 
\Phi_+(\vec p') =0
\end{equation}
Eq. (\ref{diracpp}) represents the projection of the Dirac equation onto
the positive energy states, only.
It can be solved numerically or used to obtain
a nonrelativistic reduction by means of an expansion in powers of $p/m$.\\

\vskip 0.5 truecm
\noindent
We introduce now
the Green function  for the one-body case; it  can be written as:
\begin{equation}\label{green1}
\Gamma_{(1)}= {\frac {1} { E \gamma^0 - \vec p \cdot \gamma -m}}
= [ {\frac {\Lambda^+} {E- \varepsilon} } +
                  {\frac {\Lambda^-} {E+ \varepsilon} } ]\beta 
									=G_{(1)} \cdot \beta
\end{equation}
where $E$ represents the particle energy;
in the propagator of the Feynman graphs, $E$ is replaced by $p^0$ and the singularity
of the denominator  is
avoided by means of the substitution $m \rightarrow m - i\eta$ ($\eta >0$).\\
The inverse of the one-body Green function is straightforwardly obtained in the form:

\begin{equation}\label{green1inv}
{\Gamma}_{(1)}^{-1}= \beta [ \Lambda^+ \cdot (E- \varepsilon)  +
                             \Lambda^- \cdot (E+ \varepsilon)  ]
									=\beta 	G_{(1)}^{-1}				
\end{equation}								
With standard algebra one finds:
\begin{equation}\label{deqgreen1inv}
D=-\beta{\Gamma}_{(1)}^{-1} = - G_{(1)}^{-1}	
\end{equation}
where $D$ is the one-body Dirac operator defined in Eq. (\ref{dirop}).
In consequence, the Dirac equation for an interacting particle can be 
written in the following equivalent forms:
\begin{equation}\label{diracforms}
(D+W_{(1)})|\Psi>=0,
~~|\Psi>= { G}_{(1)}  W _{(1)} |\Psi>~.
\end{equation}
The first form is the standard one, the second form has been obtained 
by means of Eq. (\ref{deqgreen1inv}); 
due to the nonlocal character  of ${ G}_{(1)}$, it must be transformed 
into an integral equation in the momentum space.

\vskip 0.5 truecm
\noindent
\textit{Two-body case}. The three-dimensional
two-body Green function can be written in a
general form (for different models) by means of the projection operators 
of the two particles:
\begin{equation}\label{g2gen}
{\Gamma}_{(2) X}= \sum_{\lambda,\xi} 
\Lambda_1^\lambda \Lambda_2^\xi ~ g_X^{\lambda \xi} \cdot \beta_1 \beta_2
 =G_{(2) X} \cdot \beta_1 \beta_2
\end{equation}
where the subscript $X$ denotes the selected model.
Specifically, for the DLE one has:
\begin{equation}\label{g2dl}
g_{D}^{\lambda\xi}={\frac {1} {E_1+E_2 -\lambda \varepsilon_1 -\xi \varepsilon_2}},
\end{equation}
note that in the CM one has $\varepsilon_1=\varepsilon_2=\varepsilon$;
consequently, for 
$(\xi, \lambda) = (+,-)~ \text{and}~ (\xi, \lambda)= (-,+)$ 
one has unphysical poles in the Green function;\\ 
for the SIE one has:
\begin{equation}\label{g2si}
\begin{split}
g_{S}^{++}={\frac {1} { E_1+E_2 - \varepsilon_1 - \varepsilon_2}} ~~ \\
g_{S}^{+-}= g_{S}^{-+}= 0 ~~~~~~~~~~~~~~~\\
g_{S}^{--}={\frac {1} {-E_1-E_2 - \varepsilon_1 - \varepsilon_2}} ~ ,
\end{split}
\end{equation}
finally, for the MWE one has:
\begin{equation}\label{g2mw}
g_{M}^{\lambda \xi}=
{\frac {\lambda \xi} {\lambda E_1+\xi E_2 - \varepsilon_1 - \varepsilon_2}}~.
\end{equation}
Note that  with respect to the SIE, in the MWE the crossed graphs 
are taken into account by means of the eikonal approximation
and give nonvanishing values
to the coefficients $ g_{M}^{+-}, ~  g_{M}^{-+} $,
while $g_M^{++}= g_S^{++},~ g_M^{--}= g_S^{--}$ .\\
The equation for the wave fuction, in all the three cases discussed here,
 is formally written as:
\begin{equation}\label{forminteq}
|\Psi>={G}_{(2) X} W_{(2)} | \Psi>~.
	\end{equation}
Given that $	{ G}_{(2) X}$ is, in any case, a nonlocal operator,
also for a local interaction $W_{(2)}$, Eq. (\ref{forminteq}) must be written 
as an integral equation in order to perform practical calculations.\\
The two-body Greeen function ${ G}_{(2) X}$ is invertible if all the
coefficients $g^{\lambda \xi}_X$ are \textit{nonvanishing}.
 This is the case of the DLE
and MWE, but not of the SIE.
The inverse has the form:
\begin{equation}\label{g2m1}
G_{(2) X}^{-1}=  \sum_{\lambda,\xi} 
\Lambda_1^\lambda \Lambda_2^\xi \cdot
{\frac {1} {g_X^{\lambda \xi}} } ~.
\end{equation}
For the DLE and MWE,
one can write the wave equation in the following general form:
\begin{equation}\label{invstand}
[-G_{(2) X}^{-1} + W_{(2)}] |\Psi>=0~.
\end{equation}
With standard calculation one finds:
\begin{equation}\label{gm2dl}
D_1 + D_2 = -G_{(2) D}^{-1}
\end{equation}
for the DLE, and:
\begin{equation}\label{gm2mw}
 D_1 S_2 + D_2 S_1 =-G_{(2) M}^{-1}	
\end{equation}
for the MWE, with the energy sign operators $S_i$ of Eq. (\ref{ensign}). 
In this way Eqs. (\ref{dirac2}) and (\ref{mw2}) are obtained.\\
For  the SIE, using the properties of the 
projectors $\Lambda_1^\lambda$, $\Lambda_2^\xi$,
with some algebra one can write:
\begin{equation}\label{eqsi}
\begin{split}
[D_1+D_2 + (\Lambda_1^+ \Lambda_2^- - \Lambda_1^- \Lambda_2^-)W_{(2)}] | \Psi>=0~~\\
\Lambda_1^+ \Lambda_2^- |\Psi>=  \Lambda_1^- \Lambda_2^+ |\Psi>=0 ~.
\end{split}
\end{equation}
Note that the interaction term is multiplied from the left by the \textit{nonlocal}
projection operators and that one has to require the second line conditions for the
\textit{ket} $|\Psi>$.
We observe that in the DLE both the  noninteracting and interaction term
are of local form; in the MWE, the noninteracting term is nonlocal but
the interaction term is local;  the SIE is globally nonlocal due to 
the projection operators that multiply the interaction term.

\noindent
Eq. (\ref{invstand}) can be also written as a coupled equation for the
 positive and negative energy components of $|\Psi>$.
These components are defined, analogously to the one-body case of Eq.(\ref{psidirpm}),
 by means of the following equation:
\begin{equation}\label{psi2dec}
|\Psi>= \sum_{\lambda, \xi}  
u_{1,\lambda} u_{2,\xi}
 \cdot |\Phi_{\lambda \xi}> ~.
\end{equation}
We also introduce 
the following projections for the interaction operator:
\begin{equation}\label{projint}
W_{(2)}^{\lambda \xi, \eta \rho } =
u_{1, \lambda}^\dag  u_{2, \xi}^\dag  W_{(2)}
u_{1, \eta}        u_{2, \rho} ~.
\end{equation} 
In this way the wave equation can be written in the form:
\begin{equation}\label{eqcomps}
\sum_{\eta, \rho} \left [ -{\frac {1} {g_X^{\lambda \xi}} } 
\cdot \delta_{\lambda \eta}  \delta_{\xi  \rho} +
W_{(2)}^{\lambda \xi, \eta \rho } 
\right ]|\Phi_{\eta \rho  }> =0~.
\end{equation}
In the case of the SI equation, one has 
$|\Phi_{+ -}> =|\Phi_{- +}>=0$.\\
Note that $g_X^{++}$ has the same form for all the models. In consequence,
if one considers only the projections onto the $++$ states
(disregarding all the other components of the wave function),
 the wave equation takes the form:
\begin{equation}\label{eqplusplus}
 \left[\varepsilon_1 +\varepsilon_2 -E_1 -E_2 +
W_{(2)}^{+ +,+ + } \right]|\Phi_{++}>~.
\end{equation}
In the last two Eqs. (\ref{eqcomps}), (\ref{eqplusplus})
the interaction operator has a nonlocal form. 
For this reason, one has to transform these equations into integral equations.
In the CM frame,  one has to specify the relative variables
of the bound system; 
 for equal mass particles, 
using 
Eqs. (\ref{eqmass}) - (\ref{rdv}),
one defines 
\begin{equation}\label{w2momsp}
{\cal W}_{(2)}^{\lambda \xi, \eta \rho }(\vec p, \vec p') =
<\vec p|W_{(2)}^{\lambda \xi, \eta \rho } |\vec p'>~.
\end{equation}  
One also has 
$g_X^{\lambda \xi}= g_X^{\lambda \xi}(m, E_T; \vec p)$;
in this way Eq. (\ref{eqcomps}) can be transformed into 
the following coupled  integral equations:
\begin{equation}\label{eqcompsint}
\sum_{\eta, \rho} \left [
 - {\frac {1} {g_X^{\lambda \xi}} }     \delta_{\lambda \eta}\delta_{ \xi \rho}
 \Phi_{\eta \rho}(\vec p) +
\int d^3 p' 
 {\cal W}_{(2)}^{\lambda \xi, \eta \rho}(\vec p, \vec p')
\Phi_{\eta \rho}(\vec p') \right ] =0
\end{equation}
The projection onto positive energy states $++$ of Eq. (\ref{eqplusplus})
becomes:
\begin{equation}\label{eqplusplusint}
[ 2 \varepsilon (\vec p) -E_T] \Phi_{++}(\vec p) +
\int d^3 p' {\cal W}_{(2)}^{+ +,+ + } (\vec p, \vec p ') \Phi_{++}(\vec p')=0
\end{equation}
In this equation, denoted as PESE, one can include, without inconsistencies, 
some retardation contributions.
Furthermore, this equation has been successfully used to study heavy quarkonium spectra in a relativistic model.

\vskip 0.5 truecm
\noindent
\textit{Comparison with the projection of our model}.
In our model the projection operator $K$ of Eq. (\ref{defk}) is a \textit{local},
energy dependent, operator. Note that; for an on-shell particle
\begin{equation}\label{compkq}
K(m,E=\varepsilon; \vec p, \vec \sigma)= U_+(m; \vec p, \vec \sigma) ~.
\end{equation}
Also the normalization factor of Eq. (\ref{norm1}) reduces to $M$ of Eq. (\ref{mdef}).\\
We observe that a correlated state of our model
$|\Psi_{corr}>$ contains positive \textit {and negative} energy
components.
To analyze this point, we introduce previously:

\begin{equation}\label{kproj}
\begin{split}
 U_+^\dag  \cdot  K =
{\frac {\varepsilon+E} {m+E} }~ \\
 U_-^\dag  \cdot   K =
{\frac { ( \varepsilon -E) ~ {\vec p} \cdot \vec \sigma   } {(\varepsilon+m)   
(m+E)} }~.
\end{split}
\end{equation} 
Then, the positive energy amplitude for a state $|\Psi_{corr}>$ 
is given, in the momentum space, by  following equation:
\begin{equation}\label{amppos}
\begin{split}
< \vec p; +|\Psi_{corr}>= N_{(1)}\cdot M \cdot K^\dag \cdot 
U_+
\cdot \Phi(\vec p)=\\
=N_{(1)}\cdot \left [
{ {\frac {\varepsilon+m} {2 \varepsilon} }} \right ]^{1/2}
\cdot {\frac {\varepsilon+E} {m+E} }
\cdot \Phi(\vec p)~.~~~~~~~~~~~~~
\end{split}
\end{equation}
For the negative energy amplitude one has:
\begin{equation}\label{ampneg}
\begin{split}
< \vec p; -|\Psi_{corr}>= N_{(1)}\cdot M \cdot K^\dag \cdot 
U_-
\cdot \Phi(\vec p)=\\
=N_{(1)}\cdot 
{\frac  {( \varepsilon -E)    } 
{ (2 \varepsilon)^{1/2}(\varepsilon+m)^{1/2}   
(m+E)} } \vec p \cdot \vec \sigma
\cdot \Phi(\vec p)~.
\end{split}
\end{equation} 
To obtain the previous equations, Eqs. (\ref{mdef}), (\ref{kproj})
and the implicit definition of $N$ of Eq. (\ref{norm1}) have been used;
$\Phi(\vec p)$ is the two component spinor  in the momentum space.\\
We note that, in any case, our model introduces automatically some negative-energy
component in the correlated Dirac wave function.\\
Again, for $E=\varepsilon$, we have 
$< \vec p; +|\Psi_{corr}>= \Phi_+(\vec p)$ and
$< \vec p; -|\Psi_{corr}>=0 $.
In this limit our model is equivalent to the standard projection 
onto positive energy states,
represented by Eq. (\ref{diracpp}) for the one-body case 
and  by Eqs. (\ref{eqplusplus}), (\ref{eqplusplusint}),
denoted as PESE, for the two-body case.

\section{Reduction of the one-body interaction}\label{reduct1}
We generalize here the procedure for calculating te one-body reduced interaction.\\
For the one-body scalar interaction we have:
\begin{equation}\label{scal}
W_{(1)}^s=\beta \cdot   V_{(1)}^s(r) 
\end{equation}
The reduced interaction is obtained by means of the reduction operator 
$K$ of  Eq. (\ref{defk}), that is:
\begin{equation}\label{scalred}
\hat  W_{(1)}^s= K^\dag W_{(1)}^s K=
V_{(1)}^s(r) -{\frac {1} {(m+E)^2 }}
\vec \sigma \cdot \vec p       
V_{(1)}^s(r)
\vec \sigma \cdot \vec p~~.
\end{equation}
In the case of a vector interaction, for the time component we have:
\begin{equation}\label{v0}
W_{(1)}^0= \mathcal{I} \cdot V_{(1)}^0(r) 
\end{equation}
The reduced interaction is:
\begin{equation}\label{v0red}
\hat  W_{(1)}^0= K^\dag W_{(1)}^0 K=
V_{(1)}^0(r) +{\frac {1} {(m+E)^2 }}
\vec \sigma \cdot \vec p       
V_{(1)}^0(r)
\vec \sigma \cdot \vec p~~.
\end{equation}
For the 3-vector part of the interaction, we take a vector function that depends,
in general, on $\vec r$:
\begin{equation}\label{vv}
W_{(1)}^v=  \vec \alpha \cdot  \vec V_{(1)}^v(\vec r) 
\end{equation}
The corresponding reduced interaction is:
\begin{equation}\label{vvred}
\hat W_{(1)}^v={\frac {1} {(m+E) }} \cdot
\left [
(\vec \sigma \cdot  \vec V_{(1)}^v(\vec r) )(\vec \sigma \cdot \vec p  ) +
(\vec \sigma \cdot \vec p)( \vec \sigma \cdot \vec V_{(1)}^v(\vec r) )
\right ]
\end{equation}
With straightforward calculations one obtains  the following  transformation equations
that can be used to simplify the previous expressions that contains \textit{two}
Pauli matrices $\vec \sigma$:
\begin{equation}\label{simpl1}
\vec \sigma \cdot \vec p~ V(r) ~\vec \sigma \cdot \vec p =
{\frac 1 2} \{ \vec p^{~2},V(r) \} +{\frac 1 2}\nabla^2 V(r) + \vec l \cdot \vec \sigma {\frac 1 r} V'(r)
\end{equation}
with the orbital angular momentum $\vec l= \vec r \times \vec p$, and
\begin{equation}\label{simpl2}
(\vec \sigma \cdot \vec V(\vec r)) (\vec \sigma \cdot \vec p )+
(\vec \sigma \cdot \vec p )(\vec \sigma \cdot \vec V(\vec r))=
\vec p \cdot \vec V(\vec r) + \vec V(\vec r)\cdot \vec p+
\vec \sigma \cdot \vec \nabla \times \vec V(\vec r)~.
\end{equation}
\section{Reduction of the two-body interaction}\label{reduct2}
For the two-body scalar interaction, we have:
\begin{equation}\label{scal2b}
W_{(2)}^s=\beta_1 \beta_2 \cdot   V_{(2)}^s(r)~. 
\end{equation}
For the product of the time components of the vector interaction,
we have: 
\begin{equation}\label{vt2b}
 W_{(2)}^t= \mathcal{I}_1  \mathcal{I}_2 \cdot   V_{(2)}^t(r)~.
\end{equation}
In both cases the reduced interaction is obtained by using
(with a similar procedure) the operators 
$K_1$ and $K_2$;
we  summarize the results in the following way:
\begin{equation}\label{scalvtred2b}
\begin{split}
\hat  W_{(2)}^c= K_1^\dag K_2^\dag W_{(2)}^c K_2 K_1 =~~~~~~~~~~~~~~~~~~~~~~~~~~\\
V_{(2)}^c(r)+ 
            \tau^c  \cdot   \left [{\frac {1} {(m_1+E_1)^2 }}
\vec \sigma_1 \cdot \vec p_1      
V_{(2)}^c(r)
\vec \sigma_1 \cdot \vec p_1  +~~~~~~~~~~~~~~~~~~~~~~~  \right.\\ \left.
               {\frac {1} {(m_2+E_2)^2 }}
\vec \sigma_2 \cdot \vec p_2       
V_{(2)}^c(r)
\vec \sigma_2 \cdot \vec p_2  \right ] +~~~~~~~~~~~~~~~~~~~~~~~\\
+{\frac {1} { (m_1+E_1)^2 (m_2+E_2)^2 }}
(\vec \sigma_1 \cdot \vec p_1) (\vec \sigma_2 \cdot \vec p_2)
V_{(2)}^c(r)
(\vec \sigma_2 \cdot \vec p_2) (\vec \sigma_1 \cdot \vec p_1)
\end{split}
\end{equation}
where the superscript $c$ denotes  the two interactions, that is $c:s,t$;
we also introduced $\tau^c$, with $\tau^s=-1$ and $\tau^t=+1$ .
\vskip 0.5 truecm
\noindent
For the product of the spatial parts of the vector interaction, we have:
 \begin{equation}\label{vv2b}
W_{(2)}^v={\vec \alpha}_1 \cdot {\vec \alpha}_2 \cdot   V_{(2)}^v(r) 
\end{equation}
The reduction is obtained by means of the operators $K_1$ an $K_2$,
applying for the two particles the procedure used for deriving Eq. (\ref{vvred}).
The result is:
\begin{equation}\label{vectred2b}
\begin{split} 
\hat  W_{(2)}^v= K_1^\dag K_2^\dag W_{(2)}^v K_2 K_1 =~~~~~~~~~~~~~~~~~~~~~~~~~~\\
{\frac {1} { (m_1+E_1) (m_2+E_2) }}\cdot~~~~~~~~~~~~~~~~~~~~~~~~~~  \\ \left[
V_{(2)}^v(r) (\vec \sigma_2 \cdot \vec \sigma_1)
(\vec p_2 \cdot \vec \sigma_2) (\vec p_1 \cdot \vec \sigma_1)+ 
(\vec p_1 \cdot \vec \sigma_1)V_{(2)}^v(r) (\vec \sigma_2 \cdot \vec \sigma_1)
(\vec p_2 \cdot \vec \sigma_2) + \right. \\ \left.
(\vec p_2 \cdot \vec \sigma_2) (\vec \sigma_2 \cdot \vec \sigma_1)
V_{(2)}^v(r)(\vec p_1 \cdot \vec \sigma_1) +
(\vec p_1 \cdot \vec \sigma_1)(\vec p_2 \cdot \vec \sigma_2)
(\vec \sigma_2 \cdot \vec \sigma_1)V_{(2)}^v(r) \right] ~.\\
\end{split}
\end{equation}
Finally, we recall that, for the Charmonium spectrum calculation,
 the momentum operators $\vec p_1,~\vec p_2$ are given in Eq. (\ref{momentumops})
and $r = |\vec r|$, with $\vec r$ given in Eq. (\ref{rdv}).

\newpage
\begin{table*}

\caption{Comparison between the experimental values 
 of the Charmonium spectrum
and the results of the model.
The states of the spectrum  are grouped 
in the three mass intervals  $I_1$, $I_2$ and $I_3$ defined in the text.
The intervals are separated by a line. 
%
The quantum numbers  $n$, $L$, $S$ and $J$ have been introduced 
in Eq. (\ref{basisho});
they represent the principal quantum number, the orbital angular momentum, the spin 
and the total  angular momentum, respectively.
All the masses are in MeV. 
The results of the columns Theor.(A) and Theor.(B) refer to the fits A and B, 
as specified in the text.}

\begin{center}
\begin{tabular}{ccccc}
\hline
\hline \\
Name & $n^{2S+1}L_J$ & Theor.(A) & Theor.(B)  &  Experiment          \\
\hline \\
$\eta_c$    &  $1^1 S_0 $     & 2984 & 2983       & 2983.9   $\pm$  0.5   \\
$J/\psi$    &  $1^3 S_1 $     & 3096 & 3096       & 3096.9   $\pm$  0.006 \\
$\chi_{c0}$ &  $1^3 P_0 $     & 3420 & 3422       & 3414.71  $\pm$ 0.30   \\
$\chi_{c1}$ &  $1^3 P_1 $     & 3504 & 3506       & 3510.67  $\pm$ 0.05   \\
$ h_c$      &  $1^1 P_1 $     & 3519 & 3521       & 3525.38  $\pm$ 0.11   \\ 
$\chi_{c2}$ &  $1^3 P_2 $     & 3564 & 3567       & 3556.17  $\pm$ 0.07   \\
$\eta'_c$   &  $2^1 S_0 $     & 3639 & 3638       & 3637.5   $\pm$ 1.1    \\
$\psi'$     &  $2^3 S_1 $     & 3685 & 3680       & 3686.097 $\pm$ 0.025  \\
\\
\hline \\
$\psi(3770)$&     $1^3 D_1 $     & 3776 &  3765   & 3773.13  $\pm$ 0.35  \\  
$\psi(3823)$&     $1^3 D_2 $     & 3816 &  3813   & 3822.2   $\pm$ 1.2   \\
$\chi_{c1}(3872)$&$2^3 P_1 $     & 3869 &  3877   & 3871.69  $\pm$ 0.17  \\
$\chi_{c2}(3930)$&$2^3 P_2 $     & 3936 &  3936   & 3927.2   $\pm$ 2.6   \\
$\psi(4040)$&     $3^3 S_1 $     & 4035 &  4034   & 4039     $\pm$ 1     \\
\\
\hline\\
$\chi{c1}(4140)$& $3^3 P_1 $     & 4148 &  4148   & 4146.8  $\pm$ 2.4    \\
$\psi(4260)    $& $4^3 S_1 $     & 4228 &  4228   & 4230    $\pm$ 8      \\
$\chi{c1}(4274)$& $4^3 P_1 $     & 4275 &  4275   & 4274    $\pm$ 7      \\
\\
\hline
\hline \\
~\\
~\\
~\\
~\\

\end{tabular}
\end{center}
\label{tabres}
\end{table*}



\begin{table*} 
\caption{Numerical values of the parameters of the model; 
$m_q$ is fixed; $M_a$ and $M_b$  define
the mass intervals.
For the other parameters,
as explained in the text, DLE and MWE stand for the two reduced equations;
 A and B stand for the two fits that have been performed.
The reported numerical values represent the results of the fits of the free parameters. 
The parameters not given in this table are vinculated as explained in the text. }
\begin{center}
\begin{tabular}{llllll}
\hline 
\hline \\   
&                &   &  & & Units     \\ 
\hline \\
 $m_q$   &  $1275$     & &           & $ $       &  MeV   \\
 $ M_a$   &  $3700$    & &        & $ $       &  MeV   \\
 $ M_b$   &  $4080$    & &       & $ $       &  MeV   \\
\hline \\   
                 &   DLE(A)   & DLE(B)   &  MWE(A)   &   MWE(B)      &   \\ 
\hline \\   
 $\alpha_v(I_1) $&  $~1.566 $ &$~1.614 $ & $~1.574 $ &   $~1.615 $   &         \\
 $\bar V_v(I_1) $&  $~1.807 $ &$~1.803 $ & $~1.806 $ &   $~1.802 $   & GeV     \\
 $d_v(I_1)      $&  $0.2405 $ &$0.2500 $ & $0.2428 $ &   $0.2510 $   & fm      \\
 $\bar V_s(I_1) $&  $0.8270 $ &$0.8187 $ & $0.8230 $ &   $0.8177 $   & GeV     \\
 $r_s(I_1)      $&  $~1.484 $ &$~1.518 $ & $~1.488 $ &   $~1.520 $   & fm      \\
 $d_s(I_1)      $&  $0.7059 $ &$0.8800 $ & $0.7149 $ &   $0.8838 $   & fm      \\
\hline \\ 
 $\alpha_v(I_2) $&  $~1.956 $ &$~1.879 $ & $~1.962 $  &  $~1.883 $   &         \\
 $\bar V_v(I_2) $&  $~2.005 $ &$~1.854 $ & $~2.001 $  &  $~1.853 $   & GeV     \\
$r_s(I_2)      $&  $~1.905 $  &     & $~1.905 $ &     & fm  \\
\hline \\   

 %
%

\hline
\hline

\end{tabular}
\end{center}

\label{tabpar}
\end{table*}


\vskip 5.0 truecm
\begin{thebibliography}{50}  
%
\bibitem{breit1}
G. Breit, Phys. Rev. {\bf 34}, 553 (1929). 
\bibitem{breit2}
G. Breit, Phys. Rev. {\bf 36}, 383 (1930).
\bibitem{breit3}
G. Breit, Phys. Rev. {\bf 39}, 616 (1932).
\bibitem{louism}
D. J. Louis-Martinez, Mod. Phys. Lett. A {\bf 27}, 1250064 (2012).
\bibitem{garciakelkar}
 F. Garc\'ia Daza, N. G. Kelkar, M. Nowakowski, J. Phys. G: Nucl. Part. Phys. 
{\bf 39} 035103 (2012).
\bibitem{kelkarnow}
N. G. Kelkar, M. Nowakowski,  Phys.Lett. B {\bf  651}, 363 (2007).  
\bibitem{kasa}
H. Kasari, Y. Yamaguchi, Phys.Lett. B {\bf 508}, 198 (2001).
\bibitem{tsibi}
G. D. Tsibidis, Acta Phys. Pol. { \bf B 35},  2329 (2004).
\bibitem{such}
J. Sucher, Phys. Rev. Lett. {\bf 55}, 1033 (1985).
\bibitem{cra1}
H. W. Crater, R. L. Becker, C. Y. Wong, P. Van Alstine, 
Phys. Rev. D {\bf 46}, 5117 (1992).
\bibitem{cra2}
H. W. Crater, P. Van Alstine, Phys Rev. D {\bf 70},034026 (2004).
\bibitem{cra3}
H. W. Crater, J. Schiermeyer  Phys. Rev. D {\bf 82}, 094020 (2010). 
\bibitem{sim1}
 Yu. A. Simonov,   Phys. Rev. D {\bf 88}, 025028 (2013).
\bibitem{sim2}
 Yu. A. Simonov,   Phys. Rev. D {\bf 90}, 013013 (2014). 
\bibitem{sim3}
 Yu. A. Simonov,   Phys. Rev. D {\bf 91}, 065001 (2015). 
\bibitem{sabe}
E. E. Salpeter, H. A. Bethe, Phys. Rev. {\bf 84}, 1232 (1951).
\bibitem{bethes}
H. A. Bethe, E. E. Salpeter, \textit{ Quantum Mechanics of one- and 
two-electron atoms}, Springer, Berlin-Heidelberg (1957).
\bibitem{bsed}
C. Itzykson and J. B. Zuber, \textit{ Quantum Field Theory},
Mc-Graw Hill, New York-London (1985).
\bibitem{lucha}
W. Lucha, F. F. Schoberl, Phys. Rev. D {\bf 87},  016009 (2013).
\bibitem{salp}
E. E.  Salpeter, Phys. Rev. {\bf 87}, 328 (1952).
\bibitem{chang} 
C. Chang, J. Chen, Commun. Theor. Phys. {\bf 44}, 646 (2005).
\bibitem{grossa}
F. Gross, Phys. Rev. {\bf 186}, 1448 (1969).
\bibitem{grossb}
F. Gross, Phys. Rev. C {\bf 26}, 2203 (1982).
\bibitem{grossc}
 C. Savkli, F. Gross,  Phys. Rev. C {\bf 63}, 035208 (2001). 
\bibitem{cst1}
S. Leit\~o, A. Stadler, M. T. Pe\~na, E. P. Biernat,
 Phys. Rev. D {\bf 96},  074007 (2017).
\bibitem{mwa}
V. B. Mandelzweig, S. J. Wallace, Phys. Lett. B, {\bf 469}, (1987).
\bibitem{mwb}
 S. J. Wallace, V. B. Mandelzweig, Nucl. Phys. {\bf A503}, 673 (1989).
\bibitem{mwc}
 N. K. Devine, S. J. Wallace, Phys. Rev. C {\bf 51}, 3222-3231, (1995). 
\bibitem{dspa}
M. De Sanctis, D. Prosperi, Il Nuovo Cim. {\bf 104 A}, 921 (1991).
\bibitem{dspb}
M. De Sanctis, D. Prosperi, Il Nuovo Cim. {\bf 104 A}, 1845 (1991).
\bibitem{dspc}
M. De Sanctis, D. Prosperi, Few Body Syst., {\bf Suppl. 6}, 532 (1992).
\bibitem{mosh}
M. Moshinsky, A. Nikitin, Rev. Mex. F\'is., {\bf 50 Supl. 2}, 66 (2004).
\bibitem{mdsf}
D. Molina, M. De Sanctis, C. Fern\'andez-Ramirez, Phys. Rev. D {\bf 95},
094021 (2017). 
\bibitem{mdsfs}
D. Molina, M. De Sanctis, C. Fern\'andez-Ramirez, E. Santopinto,
Eur. Phys. J. C {\bf 80} 526 (2020).
\bibitem{spinsym1}
 P. Alberto, A. S. de Castro, M. Malheiro, Phys. Rev. C {\bf 87}, 031301(R) (2013).
\bibitem{spinsym2}
M. De Sanctis, Acta Phys. Pol. B {\bf 50}, 853 (2019).
\bibitem{pest}
G. Pestka, L. Syrocki, Acta Phys. Pol. B {\bf 49},  1899 (2018).
\bibitem{bhagh}
Bhaghyesh, K. B. Vijaya Kumar, Commun. Theor. Phys. {\bf 55}, 1044 (2011).
%
\bibitem{chromomds}
M. De Sanctis, Front. Phys. {\bf 7}, 25 (2019).
\bibitem{mdsquint}
M. De Sanctis, P. Quintero, Eur. Phys. J. A {\bf 46}, 213 (2010).
%
\bibitem{pdg}
M. Tanabashi \textit{et al.} (Particle Data Group), Phys. Rev. D {\bf 98},
0300001 and 2019 update.

\end{thebibliography}
\end{document}